\documentclass[pdflatex,sn-mathphys]{sn-jnl}
\usepackage{multirow}
\usepackage{array}

\jyear{2021}%

\theoremstyle{thmstyleone}%
\theoremstyle{thmstyletwo}%

\theoremstyle{thmstylethree}%

\begin{document}

\title{The Time Perception Control and Regulation in VR Environment}


\author[1]{\fnm{Zhitao} \sur{Liu}}\email{zl425uestc@gmail.com}


\author[4]{\fnm{Jinke} \sur{Shi}}\email{995921268@qq.com}

\author[4]{\fnm{Junhao} \sur{He}}\email{1224256919@qq.com}
\author[4]{\fnm{Yu} \sur{Wu}}\email{wuyusilence@qq.com}
\author*[2,3]{\fnm{Ning} \sur{Xie}}\email{seanxiening@gmail.com}
\author[4]{\fnm{Ke} \sur{Xiong}}\email{67507088@qq.com}
\author[2,3]{\fnm{Yutong} \sur{Liu}}\email{1274603699@qq.com}

\affil[1]{\orgdiv{School of Aeronautics And Astronautics}, \orgname{University of Electronic Science and
Technology of China}, \orgaddress{\street{2006 Xiyuan Avenue}, \city{Chengdu}, \postcode{611731}, \country{China}}}
\affil*[2]{\orgdiv{School of Computer Science and Engineering}, \orgname{University of Electronic Science and
Technology of China}, \orgaddress{\street{2006 Xiyuan Avenue}, \city{Chengdu}, \postcode{611731}, \country{China}}}
\affil[3]{\orgdiv{Center for Future Media}, \orgname{University of Electronic Science and
Technology of China}, \orgaddress{\street{2006 Xiyuan Avenue}, \city{Chengdu}, \postcode{611731}, \country{China}}}
\affil[4]{\orgdiv{Glasgow College}, \orgname{University of Electronic Science and
Technology of China}, \orgaddress{\street{2006 Xiyuan Avenue}, \city{Chengdu}, \postcode{611731}, \country{China}}}


\abstract{To adapt to different environments, human circadian rhythms will be constantly adjusted as the environment changes, which follows the principle of survival of the fittest. According to this principle, objective factors (such as circadian rhythms, and light intensity) can be utilized to control time perception. The subjective judgment on the estimation of elapsed time is called time perception. In the physical world, factors that can affect time perception, represented by illumination, are called the Zeitgebers. In recent years, with the development of Virtual Reality (VR) technology, effective control of zeitgebers has become possible, which is difficult to achieve in the physical world. Based on previous studies, this paper deeply explores the actual performance in VR environment of four types of time zeitgebers (music, color, cognitive load, and concentration) that have been proven to have a certain impact on time perception in the physical world. It discusses the study of the measurement of the difference between human time perception and objective escaped time in the physical world.}




\maketitle

\section{Introduction}\label{sec1}

Humans’ time perception is a distinct perception from the other five senses: sight, hearing, touch, smell, and taste\cite{Harrington1085,article6}. We cannot feel time, since it is not a three-dimensional entity, and it is generally believed that humans do not have a specific organ to feel time\cite{MECK20051,article7}. Time perception is represented by the aggregation of different cognitive senses, for example, the classic physiological model of the Pacemaker-Counter Process considers time perception as a linear system\cite{IVRY2002117}. In other words, when we get new information, our brain needs to first reorder it and then deliver it to other parts of the brain. That is the reason why we feel time flows slower when exposed to an unfamiliar environment or new information since our brain needs longer time to process\cite{article5}. To sum up, when we are under some specific situations, our brain will work harder to collect and process information. In that case, we will feel time seems to be stretched and pass slower.

\begin{figure}
    \centering
    \includegraphics[width=\textwidth]{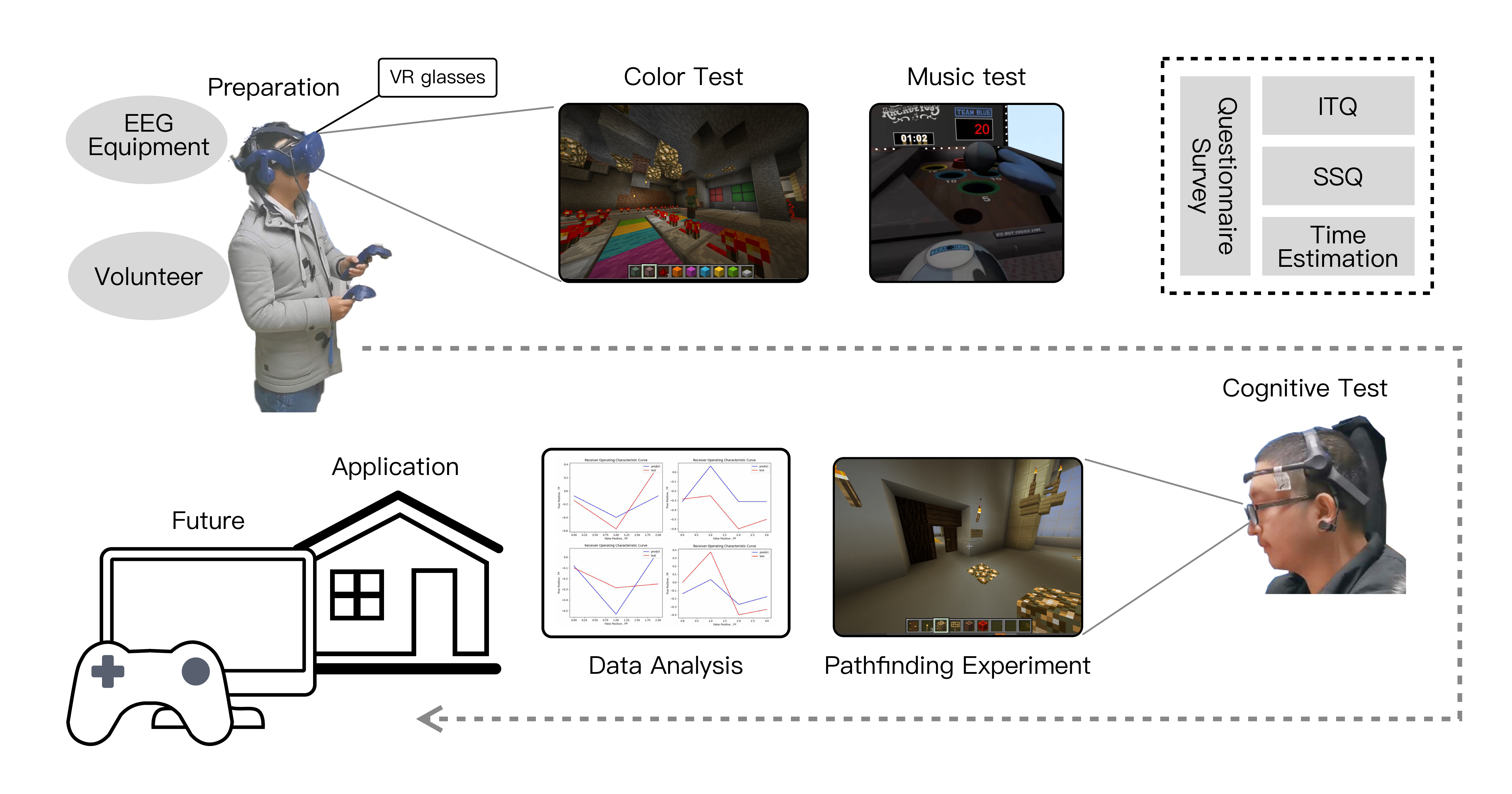}
    \caption{The system schematic. During the experiment, the participants need to wear HTC VIVE Pro Head Mounted Display (HMD). They will first be asked to familiarize themselves with the VR environment, then conduct color and music experiments. Then, the subjects were asked to wear a portable EEG device (TGAM) to conduct cognitive load and concentration experiments, and based on this experiment, the tension experiment was carried out. By collecting and analyzing data, this paper aims to figure out the impact of five zeitgebers on users' time perception in the immersive virtual environment.}
    \label{fig:my_label}
\end{figure}

The time perception can be manipulated by some objective factors since human circadian rhythms will adjust to the changes of the environment. For example, although the human biological clock can be expressed as naturally as a pre-programmed program, it can be influenced by the light-dark cycle\cite{article1}, temperature changes\cite{article0}, time of three meals, social activities\cite{10.1001/archpsyc.1988.01800340076012}, and other factors. Generally, these factors are signals from the environment that coordinate the circadian rhythm of humans with the external environment and are therefore called zeitgebers.

\section{Related Works}\label{sec2}

As early as 1965, Aschoff has proposed the basic concept of zeitgeber\cite{Aschoff1427}, and pointed out that the most important and direct zeitgeber in nature is ”light”, which has an irreplaceable influence on human health\cite{ROENNEBERG2016R432}. More Specifically, different external light will send optical signals to the biological clock by passing through the retina’s optic nerve every morning or evening\cite{ROENNEBERG2007R44}. After receiving these signals, the biological clock will regulate the corresponding circadian rhythm under the affection of this zeitgeber. This can also explain why ancient people work when the sunrise and rest when the sunset. Zeitgeber has been used in many areas of life\cite{Ehret1212,article3}, mainly in the medical field\cite{10.1093/brain/awh135}, the patient’s time perception can be changed by using drugs or hypnosis\cite{article4}.

\begin{figure*}[t]
\centering
\includegraphics[width=\linewidth]{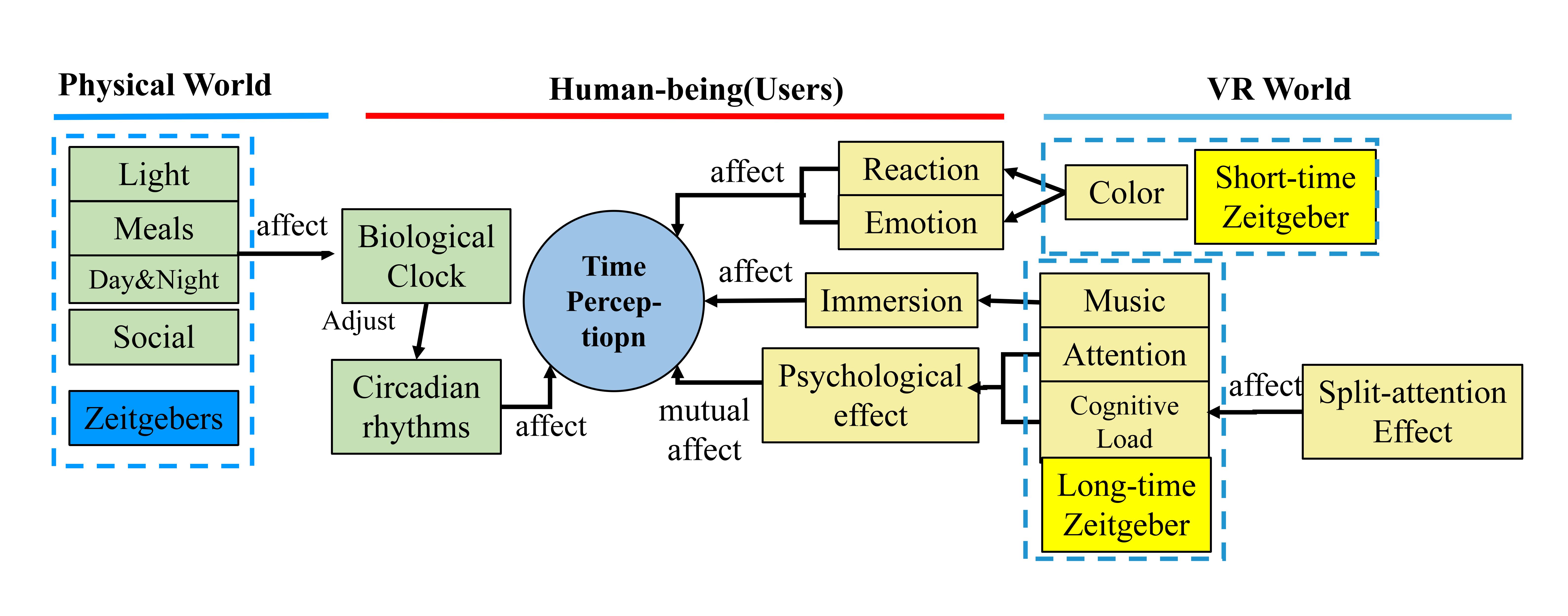}
\caption{Difference of Zeitgebers between VR Environment and Physical World}
\label{fig:fig2}
\end{figure*}

In recent years, the rapid development of VR (Virtual Reality) technology makes us start thinking about whether the zeitgeber, which has been widely used in the physical world, can also affect the virtual world. However, some researches have shown that situation in VR is different\cite{9089568}. On the other hand, the application of zeitgeber in VR has much greater potential than in the physical world. Not only because many IVEs (Immersive Virtual Environments) support near-natural audio-visual stimuli(\autoref{fig:fig2}), similar to those we generate in the physical world, but also most physical parameters can be modified in VR environment\cite{7383328}. The basis and rationale for the classification in (\autoref{fig:fig2}) will be explained in later sections

Our work is inspired by \cite{9089568}, in which Liao et al. study various zeitgebers for their experiments. For example, Liao divides the zeitgeber into auditory and visual, while this paper divides them into short-time and long-time. As the name implies, the former acts on a few minutes, or even a few seconds of time perception. It is mostly used to affect the user’s instantaneous reaction speed. The latter acts on subjective time estimation over a period of time and have a significant impact on the overall experience of a given environment. There are two reasons for this design. Firstly, the visual zeitgeber has not been found to have significant potential in the VR experiment, and its influence on time perception is not stable. Secondly, emotional stimulation for a long time (2-6 min) will decrease the arousal degree compared with the stimulation with a short time, and prolonged-time usually has greater effects on user' time perception (especially in VR)\cite{DROITVOLET2020103170}.

Addtionally, immersion is not considered as a variable which has effect on time perception in this paper. The first reason is that conclusion about immersion is usually not convincing enough due to the lack of a complete set of methods for quantifying immersion. The second reason is that some researches suggest that immersion and time perception may be separated\cite{2020The} and do not seem to change the time perception\cite{Nordin2014Immersion}.

We hope to find out the effect of several important zeitgebers on time perception, rather than simply determine whether it is related to the time perception. Therefore, we selected four representative and significant zeitgeber to study.

\section{Theoretical Basis of Zeitgebers}
\subsection{Audiovisual Zeitgebers: Color and Music}
When we were investigating different zeitgebers, we noticed a unique factor called color. Color can regulate time perception by affecting human mood and instantaneous reaction time, that's a short time. The most typical example is the red warning lights used by ambulances to alert surrounding people in an emergency. Indeed, in the physical world, some applications using color as a long-time zeitgeber, such as the use of different colors to regulate the mood of guests during dining, to achieve a higher turnover rate\cite{YILDIRIM20073233,TANTANATEWIN2018124}. To sum up, color is an extremely important zeitgeber in the physical world, and one of the few factors that can affect short-term time estimation.

However, in a VR environment, we only focus on the short-time effects of color. It is generally believed ability to react to changes in short time intervals may share the same mechanism as time perception, which means we could quantify human’s time perception by measuring their fast-paced movement perception of the physical world, and we assume that the influence of the different color for time perception in short interval may have obvious difference\cite {https://doi.org/10.1111/1469-7610.00043}.

Whether in the physical\cite{10.3389/fnhum.2015.00212} or the virtual world\cite{article10}, red is the color that has the best performance in attracting people’s attention. It is commonly seen in many dangerous marks or advertising. In the physical world, many studies have explored the effect of color on time perception. One of the typical researches is, red is better than blue to make participants concentrate under short visual stimulation conditions\cite{Thnes2018ColorAT,Katsuura2007EffectsOM}. This type of research is generally conducted by presenting participants with a computer screen and asking them to compare the duration of two different colors\cite{Shibasaki2014TheCR}. These researches lack further exploration on the practical application of result.  We proposed a more realistic application of the experiment to verify the impact of different colors on time perception.

Since color may have a significant effect as a short-time zeitgeber in VR, what is the typical long-term zeitgeber corresponding to it? Corresponding to it, we chose music as long-time zeitgeber, not only because they are both indispensable and important parts of the VR environment, but also because the current research on the effect of music on the time perception has not been very conclusive.

Specifically, as an indispensable element in the VR environment, background music plays a vital role, it can greatly improve the gaming experience and sense of immersion of VR games\cite{Gormanley2017AudioII}. By arousing specific emotions, music can affect human time perception indirectly\cite{10.3389/fpsyg.2013.00417}. However, most current researches focus on exploring the influence of music on players’ concentration, immersion, and other factors\cite{Gallacher2017GameA}. Their experiment conclusions are generally based on analyzing participant heart rate\cite{9033756} and EEG data\cite{BHATTI2016267}, few consider the participant’s sense of time. Also, these experiments are too focused on the experiment itself, for example, the participant is only allowed to listen to music during the experiment. The degree to which music in a game affects the user experience may be related to whether the player is noticed of the music, and in some cases, the music may be ignored when the player's sense is being affected by other factors of the game\cite{2020PotentialDisconnect}. Therefore, we should not only select suitable background music, but also select a suitable testing environment that will not interfere with the background music.

\subsection{Psychological Cognitive Effect Zeitgebers: Cognitive Load and Attention}
Regrettably, previous studies on color and music have no definitive conclusion since the influence of some environment variables, particularly immersion. Although our experiment design has learned from and improved this point, it is still difficult to make a qualitative leap in conclusion. In that case, we wanted to find zeitgebers that had a vital effect on the time perception, and it must possess realistic meaning. After much deliberation, we settled on cognitive load and attention. Both of them are concepts put forward by cognitive psychology. Essentially, they regulate the sense of time by influencing the psychological cognitive effect\cite{article8}. Especially in the circumstances which have obvious individual differences reflected more significant effects\cite{ZELANTI2011143}.

For former, even in some studies\cite{9089568}, cognitive load is considered as a parallel factor to zeitgeber, but we take it as an important zeitgeber\cite{PILLAI2020106389}. Related research has also been applied to the safe driving system, using professional eye tracker and external conditions to judge the intensity of cognitive load, and then infer the reaction time to cope with different cognitive load in real road conditions\cite{PILLAI2020106389}. We designed an experiment to investigate users' estimates of time elapsed in a VR environment under different cognitive loads. In part of researches, cognitive load is one of the most important zeitgebers, which has a significant impact on time estimation in reality\cite{ISRAEL2021101633}. People under high cognitive load are more likely to be impulsive than those who under low cognitive load, that is, people under high cognitive load feel time flows slower. In general, cognitive load is the most important part of this paper, and the effect of cognitive load in virtual reality environment still need to explore, since we strongly explore the effect of cognitive load on time perception.

For latter, Attention is a zeitgeber that we realized during experiment, we found through the questionnaire survey that in many cases, data could not explain the phenomenon, because the individual differences of users were too obvious. Therefore, in the second half of our research, we supplemented attention data collected using EEG devices as an important attempt to quantify individual differences. 

Attention is one of the most important Zeitgebers which partly reflect individual differences and state of the participant when testing. Its impact on the time perception is usually considered to by influencing Perceived Duration of Intervals\cite{article8}, which has an important position in many physiological models of time perception\cite{article7,BLOCK2014129}. 

\section{Methods for Quantifying Zeitgebers}
\subsection{Maximum Reaction Interval of Color}
In this paper, a user-oriented HSV(Hue, Saturation, Value) color model was adopted to explore the color influence on time. As the HSV value is high, the color is bright and white, and it has been proved is proportional to the color sensitivity\cite{8797752,doi:10.3109/07420529709001449}. To reflect the impact of different color systems on the time perception, rather than making them more eye-catching using brightness(value), we set the same value in HSV for test colors(\autoref{tab:tab1}).

\begin{table}[h]
\centering
\caption{Parameter Selection of Test Color in HSV Model}
\label{tab:tab1}
\begin{tabular}{cccc}
\toprule
           &\textbf{Hue}  & \textbf{Saturation} & \textbf{Value} \\
           \midrule
\textbf{CadetBlue}  & 0.51 & 0.41       & 0.93  \\
\textbf{Chartreuse} & 0.25 & 0.98       & 0.93  \\
\textbf{Yellow}     & 0.17 & 0.98       & 0.93  \\
\textbf{Grey}       & 0.00 & 0.00       & 0.93 \\\bottomrule
\end{tabular}
\end{table}

\begin{figure}[h]
\centering
\includegraphics[width=.8\linewidth]{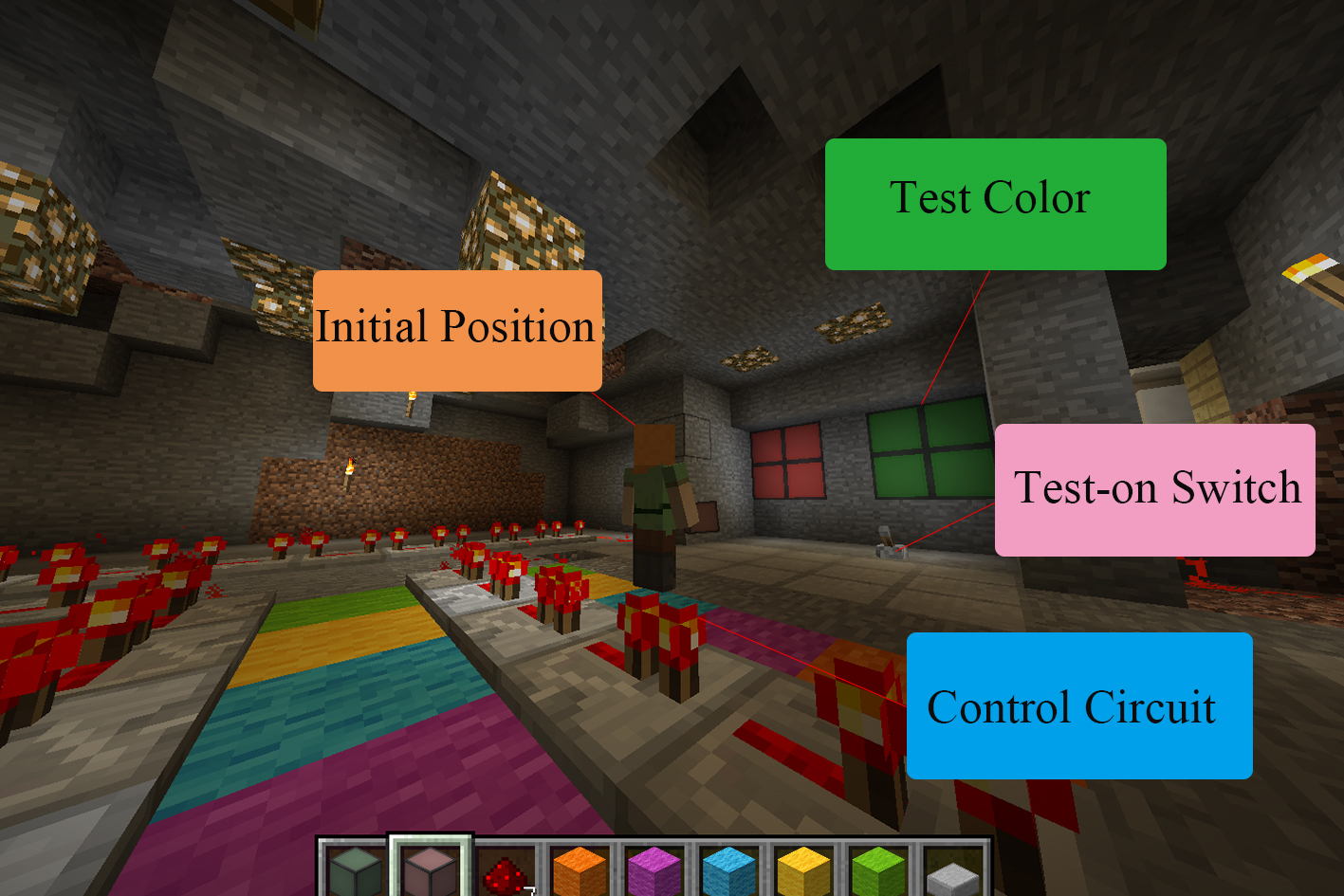}
\caption{Schematic Diagram of Experiment Environment for short-time zeitgeber (Color)}
\label{fig:fig3}
\end{figure}

To ensure that the experimental results have practical application significance, rather than simply reproducing the previous experiments in the VR environment again, our experimental design is based on the function of color in VR games. This is because how to draw the player’s attention through color as a reminder in a highly complex environment with multiple colors in VR is an important issue. Also, whether a VR scene can be modified to enhance player immersion by changing the tone of the scene, and has profound application potential. In this experiment, the \emph{Minecraft}$^{TM}$ (Mojang Studios, 2009)$^{i}$ environment was set up in VR. To simulate a game session, we built a puzzle-type game in a dark basement. As shown in \autoref{fig:fig3}, in the center of the basement there is a screen with light blocks of different colors. The left side of the screen is a red light block set as the control group, and the right one is test colors for the experimental group. This figure shows that subjects only face walls and test light blocks, while behind is the circuit system built using the game module. As a result, the subjects were not disturbed by colors other than the walls. While the walls simulate the ambient color of the real application, we set them relatively dim to maximize the impact of the test light blocks.

The experiment design refers to the time discrimination principles\cite {https://doi.org/10.1111/1469-7610.00043}, namely, test the participants’ time perception by conducting them to judge the order in which the lights go out. Each participant will undergo 4 tests for 4 test colors. In the initial condition with both lights on, the red light was fixed for 1,000 ms after the participants pulled the switch. At the beginning, the test light was lit for 700ms, then the participants need to decide which light went out first. If they can still choose the right light block, extending the test light block lighting time by 10ms and then ask them to make a judgment again, until the participants could no longer tell the difference between the two light blocks. Their sensation about the difference were recorded and classified into three categories: clear, vague, and invisible. It is noted that when the critical scintillation frequency is reached at 70Hz, scintillation cannot be perceived by the human eyes, but the light modulation can still be sensed slightly\cite{KENT2020107060}. Therefore, we estimated that the participants could not extinguish the time difference when the time interval between two lights was less than 15ms and chose 10ms as the interval to continuously reduce the time interval.

Based on previous studies and inferences, we make the following hypothesis: for different test color, the results of different participants will be significantly different.

The specific logic is shown in Algorithm 1, firstly we define \emph{s} as sensitivity of participants in the current time interval. \emph{s} = 0,2,3 represents invisible, vague and clear respectively. Therefore, the ideal situation is that \emph{s} is always equal to 3 and the graph is a straight line parallel to the horizontal axis. Then, since it is much easier to change from clear to vagueness (3 to 2) than from vagueness to invisibility (2 to 0), we double the weight of the latter. 
 
\begin{algorithm}[h]
\caption{Color Data Analysis Logic}
\label{alog:a1}
\begin{algorithmic}[1]
\State \textbf{define:}
\State \qquad x, y \textbf{as} jumped point position
\State \qquad color sensitivity \textbf{as} s
\State \textbf{set} weight:
\State \qquad clear to vagueness \textbf{as} $s_{1}$ = 1
\State \qquad vagueness to invisibility \textbf{as} $s_{2}$ = 2
\State Area = $s_{1}$ $\times$ x + $s_{2}$ $\times$ y = x + 2y
\State \textbf{run} Isolation Forest \textbf{to} remove outliers
\State \textbf{cal} $s_{total}$ \textbf{and} $Area_{total}$
\State \textbf{regularized} $s_{total}$:
\State \qquad \textbf{insert} average filite
\State \qquad \textbf{obtain} slope \textbf{of} changing trend
\State \textbf{synthesize} $Area_{total}$ \textbf{and} slope \textbf{for} conclusion
\end{algorithmic}
\end{algorithm}

\begin{figure}[h]
\centering
\includegraphics[width=0.9\linewidth]{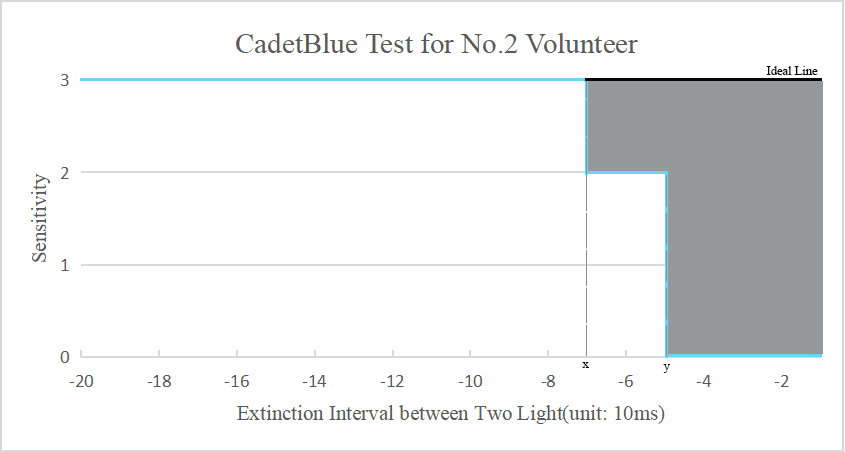}
\caption{A Sample Set Data for Color Experiment}
\label{fig:fig4}
\end{figure}

Secondly, in order to quantify the user’s sensitivity to color, we define a variable: Maximum Reaction Interval. We use MRI to represent it in rest text. MRI is inversely proportional to the sensitivity of the color, which is the shaded part in \autoref{fig:fig4}. This figure is one group of the test data, and the test light color is CadetBlue. Node \emph{x} is where the participants' sensitivity jumped from \emph{s}=3 to \emph{s}=2, and node \emph{y} for \emph{s}=2 to \emph{s}=0(from the most left side and take the absolute value). In that case, the larger the value of x and y is, the earlier the time interval between two lights can be observed. Therefore, Area is inversely proportional to the participant’s sensitivity to color, which is equivalent to a deducted score. In other words, the larger the Area of the grey part (\emph{x}+2\emph{y}) is, the larger the time interval is needed to be observed, and the less sensitive the participants are to this color. Before analyzing the data, we used the Isolation Forest method to screen for outliers and removed three sets of data that were clearly outside the reasonable range \autoref{tab:tab2}).

It can be clearly seen from the \autoref{tab:tab2} that the area in grey is significantly higher than the other three in terms of average or total value. Indeed, from common sense and previous studies, grey is almost one of the least obvious colors, which can be used to delay instantaneous response and prolong subjective time perception.

\begin{table}[h]
\centering
\caption{Relationship between the x, y and MRI ( x+2y )}
\label{tab:tab2}
\scalebox{1}{
\begin{tabular}{ccccccccccccc}
\toprule
      & \multicolumn{3}{c}{\textbf{Cadet Blue}} & \multicolumn{3}{c}{\textbf{Chartreuse}} & \multicolumn{3}{c}{\textbf{Yellow}} & \multicolumn{3}{c}{\textbf{Grey}} \\
\midrule
      & MRI     & X      & Y    & MRI      & X     & Y     & MRI      & X      & Y     & MRI     & X     & Y     \\\midrule
\#1     & 7        & 3      & 2    & 6         & 2     & 2     & 8         & 4      & 2     & 11       & 5     & 3     \\
\#2     & 17       & 7      & 5    & 11        & 5     & 3     & 11        & 5      & 3     & 16       & 6     & 5     \\
\#3     & 11       & 5      & 3    & 16        & 6     & 5     & 7         & 3      & 2     & 10       & 4     & 3     \\
\#4     & 12       & 6      & 3    & 10        & 4     & 3     & 10        & 4      & 3     & 14       & 5     & 5     \\
\#5     & 14       & 6      & 4    & 17        & 7     & 5     & 9         & 5      & 2     & 11       & 5     & 3     \\
\#6     & 6        & 4      & 1    & \textbf{1}           & \textbf{1}      & \textbf{0}       & \textbf{2}           & \textbf{2}       & \textbf{0}       & 10       & 4     & 3     \\
\#7     & 8        & 4      & 2    & 7         & 3     & 2     & 7         & 3      & 2     & 10       & 4     & 3     \\
\#8     & \textbf{22}       & \textbf{10}       & \textbf{6}      & 7         & 3     & 2     & 13        & 5      & 4     & 16       & 6     & 5     \\\midrule
Total & 75       &        &      & 74        &       &       & 65        &        &       & 99       &       & \\ 
\bottomrule
\end{tabular}}
\end{table}

\subsection{Pace and Matching Degree of Music}
To simulate this game section, we selected four scenes provided by SteamVR, all of which are game themes suitable for players to explore. For example, one of the scenes is taken from the VR game \emph{Half-Life: AlyxTM}$^{
TM}$ (Valve, 2020)$^{ii}$. In the selection of experimental music, to avoid the influence of language on the participant, and to make the experiment music play the role of background music to the greatest extent, all experiment music selected from the original soundtrack of the movies and games. Also, all of them are edited into the audio clip of the same length based on not affecting the musicality. At the same time, to avoid the subjective influence brought by the music selection, each scene contains two sets of experimental music, one has a faster pace, the other one has a slower pace.

For each experiment scene, four groups of music will be tested: two of them are experimental music, as the experimental group, one is the light music control group, and the other is the no music control group. The testing time of four experiments is controlled at the same length, nor each experiment scene, four groups of music will be tested: two of them are experimental music, as the experimental group, one is the light music control group, and the other is the no music control group. The testing time of four experiments is controlled at the same length, ninety seconds, and the participants are not informed of the testing time in advance. After the music starts playing, the participant is allowed to move freely in the scene and explore; when the music stop playing, the participant needs to answer his subjective length of time. Also, at the start of the experiment, the participant needs to explore the scene for one minute and this time length will be informed. This is to let participant has a basic concept of the speed of the time and avoid large differences in the experimental results of different participants.

The selection of the four scenes and their test music is shown in \autoref{tab:tab3}. The second scene, Stormwind, for example, is taken from the famous MMORPG (Massively Multiplayer Online Role-playing Game) \emph{World of Warcraft}$^{
TM}$ (Blizzard Entertainment, 2004)$^{iii}$, the main City of the Alliance in World of Warcraft. In this experiment, we chose Stormwind (City Theme) and Legend of Zelda Main Theme Medley as the test music of the two experimental groups. The latter comes from the well-known Japanese RPG(Role-playing Game), The \emph{Legend of Zelda: Breath of the Wild}$^{
TM}$(Nintendo, 2017)$^{iv}$ game soundtrack. The two distinct test music groups were chosen as the experimental group to study the matching degree, which will be explained in detail later.

\begin{table}[h]
\caption{Music Experiment Environment and Background Music Selection}
\label{tab:tab3}
\centering
\scalebox{0.9}{
\begin{tabular}{ccccc}
\toprule
\multirow{2}{*}{Scene} 
                    & \multicolumn{2}{c}{Experimental Group1}  & \multicolumn{2}{c}{Experimental Group1}     \\ \cline{2-5} 
                    & BGM              & Source   & BGM              & Source        \\ \midrule
Universe                  & Cornfield Chase & Interstellar & Life            & Prometheus   \\
Stormwind                 & Stormwind       & World of Warcraft & Road to Camelot & Legend of Zelda   \\
Half-Life                 & Sound Effect            & Half-Life  & Drone Attack    & Oblivion                 \\
Scenery                  & Panorama        & Light Music  & Forest              & Light Music              \\ \bottomrule
\toprule
\multirow{2}{*}{Scene}
                    & \multicolumn{2}{c}{Control Group1} & \multicolumn{2}{c}{Control Group2} \\ \cline{2-5}
                    & \multicolumn{2}{c}{BGM} & \multicolumn{2}{c}{BGM} \\ \midrule
                Universe &\multicolumn{2}{c}{\multirow{4}{*}{The Floating Cloud}} & \multicolumn{2}{c}{\multirow{4}{*}{None}} \\
                Stormwind  \\
                Half-Life \\
                Scenery \\
\bottomrule
\end{tabular}}
\end{table}

\subsection{Split-attention Effect of Cognitive Load}
The design of this experiment refers to the map exploration part of the open-world game, how players can reach the designated position faster after receiving the task assigned by NPC(non-player character). Unlike PC or console platforms, VR devices are not easy to wear, so players cannot search strategy online while playing. Therefore, a VR game developer needs to set clearer instructions to help the player finish the task when the game scene contains complex elements.  

We designed a pathfinding mission, set up an MC environment like the color experiment, and designed three different routes. After a certain amount of prediction tests, it was proved that the three routes have the almost same complexity. This is for preventing that experience the same route more times will bring different proficiency, which may lead to the deviation of the result. The subjects were tested three times and followed a given guide. High, medium, and low, three cognitive loads were randomly assigned to the three routes, which corresponded to the text guide, graphic guide, and video guide, respectively.

For the volume of participants, G*power was used for a priori estimation to determine how much sample size we needed to draw a strong conclusion. We chose the single-tail test. Although the single-tailed test cannot reach the significance level of the two-tailed test, it is more sensitive to exploring whether there is a strong correlation between the two parameters. Besides, the experiment has a strong directional expectation. In other words, according to previous studies(higher CL brings slower time), we are sure that the low-probability event on one side will not happen. Generally, $\alpha$ was selected as 0.05, statistical test power (1-$\beta$) was selected as 0.8, and the effect size was selected as 0.8. For the simulation results obtained, 32 groups of data are needed. In that case, we selected 12 volunteers to carry out 36 groups of experiments. All the participants were exposed to the VR environment for the first time.

Cognitive load theory is referred to explain why different types of guidance differ in cognitive load. Specifically, the limited processing capacity of the human brain severely limits the acquisition of complex cognitive skills. The famous cognitive psychologist John Sweller first proposed this theory in 1988, its theoretical system is gradually improved. The present cognitive load theory provides Structuring Instruction to promote teaching efficiency, such as the famous goal-free Effect\cite{2010Cognitive}. Our experiment uses the split-attention effect\cite{Mayer1998A}, which means replacing multiple sources of information with one integrated source can reduce cognitive load\cite{2009Explaining}. This effect was why we set different cognitive loads by text, graphic, and video guidance. Also, this effect mainly influences novices, which exactly corresponds to the participants who are exposed to MC (VR platform) for the first time.

\section{Experimental Results and Analysis}
\subsection{Effect of Color on Reaction Speed}
\begin{figure}[h]
\centering
\includegraphics[width=.8\linewidth]{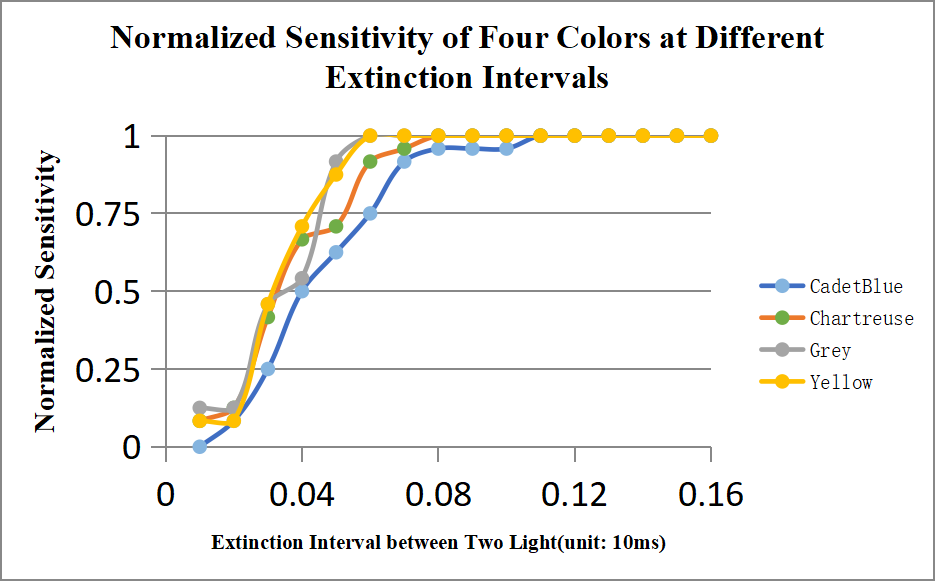}
\caption{Normalized Sensitivity of Four Colors at Different Extinction Intervals}
\label{fig:fig5}
\end{figure}

In the previous section, we use observational method directly to compare MRI to determine color sensitivity. However, the conclusion obtained from this aspect has some limitations, because we did not consider the changing trends of \emph{s} of each participant. Therefore, finally, according to the above data, the \emph{s} of all participants in each time interval are regularized and averaged, and the average filter is used to remove the high-frequency information in the curve and obtain the changing trend of the sensitivity of the 4 colors under different time intervals(\autoref{fig:fig5}). The ideal situation in this case is that all participants have a sensitivity of 3, which is 1 after regularizing.

For a highly sensitive color, the regularized mean value of \emph{s} should change more smoothly in this case. This is because the smaller the area corresponding to the color with higher sensitivity, the less drastic change in the overall mean value. Judging from the slope of each curve, cadet blue and grey are still the most and least vivid colors, respectively. It should be noted that although the blue in \autoref{fig:fig5} reaches its peak at the latest, it only proves that this color is too early from the transition point of clear and vague perception, and this shows no obvious connection with sensitivity. 

Based on the above analysis, our conclusion is: in the VR environment, the user's sensitivity to the four colors is cadet blue > chartreuse $\geq$ yellow > grey. The use of highly sensitive colors is conducive to enhancing the player's instant response and shortening the time estimation. Indeed, in the questionnaire survey after the experiment, we found that the experiment time of sensitive color is estimated to be shorter.

As mentioned above, we chose these colors to avoid the influence of the value in the HSV model, namely the brightness, on the experimental results. In this case, according to MRI and weighted by the slope(\autoref{fig:fig5}) of the normalized sensitivity curve (CadetBlue: 0.40, Chartreuse: 0.51, yellow: 0.80, grey: 0.75), we can get the following color sensitivity spectrum with the most sensitive color red as the reference. The closer the color is to red, the more conducive to enhancing the reaction speed. The faster the reaction, the faster the subjective time, the shorter the estimated time.

\begin{figure}[h]
\centering
\includegraphics[width=\linewidth]{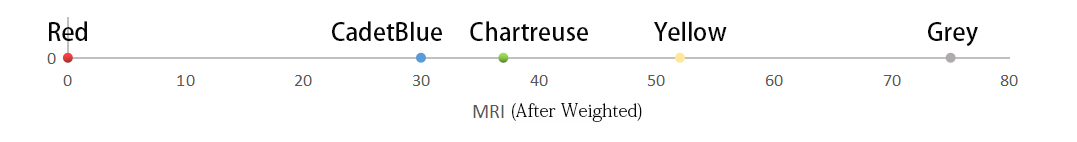}
\caption{Color Sensitivity Spectrum (take red as reference)}
\label{fig:fig6}
\end{figure}

\subsection{Limited Influence of Music}
In order to explore whether BPM will affect users' time perception, a normal distribution test is conducted on Music Data. First, we treated control group 2 as a comparison and did not add it in test process. After that, all the data in the same scene were tested, and then the data of same music type (same test group) were tested. This is because we may conduct comparative analysis on the scene or music in the experimental analysis part, so we need to confirm that the data are normal from both directions. Specifically, each scene contains 32 groups of experimental data, and each test group(music type) contains 24 groups of data. TD (Time Difference) is used as the test target, and Shapiro-Wilk test is used as the test method. The reason for using Shapiro-Wilk test is that each group of data is not more than 32 groups, which is a small sample test. When the output p-value is greater than or equal to 0.05, the data can be considered as normal distribution \cite{2011Power}. The specific test results are shown in \autoref{tab:tabx}(EG: Experimental Group, CG: Control Group). It can be seen that the data are normal in all aspects.

\begin{table}[h]
\centering
\caption{The Results of Using Shapiro-Wilk to Verify Normal Distribution of Music Experimental Data}
\label{tab:tabx}
\scalebox{1.0}{
\begin{tabular}{cccccccc}
\hline
 & \multicolumn{3}{c}{\textbf{Music Comparison}} & \multicolumn{4}{c}{\textbf{Scene Comparison}} \\ \cline{2-8} 
 & EG1 & EG2 & CG1 & Universe & Stromwind & Half-Life & Scenery \\ \hline
tests statistic & 2.18 & 2.87 & 4.03 & 2.53 & 2.93 & 1.67 & 2.63 \\
p-value & 0.34 & 0.24 & 0.13 & 0.28 & 0.23 & 0.43 & 0.28 \\ \hline
\end{tabular}}
\end{table}

To explore whether the pace of music affects the player's perception of time, we performed a Kendall tau test, a non-parameter hypothesis test, to determine their relevance \cite{abdi2007kendall}. To calculate the Kendall rank correlation coefficient tau, we ranked the pace of music according to its BPM and the time difference according to its value, specific logic can be seen in Algorithm 2.

\begin{table}[h]
\caption{Experimental Data to explore the influence of music on time perception in one scene(TD:Time Difference)}
\label{tab:tab4}
\centering
\scalebox{0.9}{
\begin{tabular}{ccccc}
\toprule
 & \textbf{TD}                      & \multicolumn{1}{l}{\textbf{Results Level}} & \multicolumn{1}{l}{\textbf{Music Level}} & \multicolumn{1}{l}{\textbf{Concordant/Discordant}} \\ \midrule
\#1            & -33.33\% & 2                                 & 1                               & 12/0                                    \\
\#2            & -22.22\%                     & 2                                 & 1                               & 12/0                                    \\
\#3            & -22.22\% & 2                                 & 1                               & 12/0                                    \\
\#4            & -11.11\% & 3                                 & 1                               & 3/4                                    \\
\#5            & -33.33\% & 2                                 & 1                               & 12/0                                    \\
\#6            & -44.44\%                     & 1                                 & 1                               & 1/0                                    \\
\#7            & -22.22\%                     & 4                                 & 1                               & 0/9                                    \\
\#8            & -33.33\%                     & 2                                 & 1                               & 12/0                                    \\
\midrule

\#1            & 0.00\%                       & 3                                 & 2                               & 1/2                                    \\
\#2           & -11.11\%                     & 3                                 & 2                               & 1/2                                    \\
\#3           & 33.33\%                      & 4                                 & 2                               & 0/7                                    \\
\#4           & 0.00\%                       & 3                                 & 2                               & 1/2                                    \\
\#5           & 33.33\%                      & 4                                 & 2                               & 0/7                                    \\
\#6           & -38.89\%                     & 2                                 & 2                               & 6/0                                    \\
\#7           & 0.00\%                       & 3                                 & 2                               & 1/2                                    \\
\#8           & -33.33\%                     & 2                                 & 2                               & 6/0                                    \\
\midrule

\#1           & -16.67\%                     & 3                                 & 3                               & 0/0                \\
\#2           & -16.67\%.  & 3                                 & 3                               & 0/0                 \\
\#3           & 11.11\%                      & 3                                 & 3                               & 0/0                 \\
\#4           & -22.22\%                     & 2                                 & 3                               & 0/0               \\
\#5           & 33.33\%                      & 4                                 & 3                               & 0/0                 \\
\#6           & -27.78\%                     & 2                                 & 3                               & 0/0                \\
\#7           & 11.11\%                      & 3                                 & 3                               & 0/0               \\
\#8           & -16.67\% & 3                                 & 3                               &  0/0                 \\
\midrule
Total        &                             & \multicolumn{1}{l}{}              & \multicolumn{1}{l}{}            & 95/35                                   \\ 
\bottomrule
\end{tabular}}
\end{table}

After determining that the data were normally distributed, we chose the Kendall Tau Test, a nonparametric hypothesis test, to determine their correlation. Since we conducted a total of 96 groups of tests in four scenarios, the experimental results are too much. Here, we only show the results of one experiment scenario(\autoref{tab:tab4}). 

Firstly, TD(time difference) is the difference between the subjective estimate time and the objective elapsed time in percent. For TD of -40\% and above, the subjective time estimation is 40\% shorter than the objective time, classification method is shown in algorithm 2. Then, we divided the music into five levels according to the rhythm. In this example, music with levels 1, 2, and 3 have BPM(beat per minute) of 87.71, 101.55, and 115.66, respectively. 

Sencondly, we conducted statistics on the data to find the Kendall rank correlation coefficient(\autoref{eq:music_eq}) to determine whether there was a strong correlation between the sequences, and Concordant and Discordant represent C and D in equation. Correlation coefficient obtained is 0.3125, for other three groups are 0.0625,0.0360 and -0.2656. It indicates that the pace does not significantly affect the time perception of testers.

\begin{equation}
\label{eq:music_eq}
    {\tau-c}=\frac{C-D}{\frac{1}{2}N^{2}\frac{M-1}{M}}
\end{equation}

Additionally, we also explored the impact of the degree of matching between music and game scene on time cognition. Unfortunately, due to individual differences, this parameter has not much reference value. Take Stormwind scenes as an example. In theory, most participants should think that the former soundtrack(\autoref{tab:tab3} experimental group1) matches the experimental scene better. However, the results were surprising. Only half of the participants thought that the Stormwind was more compatible with the scene, and a considerable part of participants thought that it was not even as good as the light music. 

\begin{algorithm}[h]
\caption{Music Data Analysis Logic}
\label{alog:a2}
\begin{algorithmic}[1]
\State \textbf{transfor} TD(\%) to RL:
\State \qquad less than -40 \textbf{as} RL = 1
\State \qquad -40 to -20 \textbf{as} RL = 2
\State \qquad -20 to 20 \textbf{as} RL = 3
\State \qquad 20 to 40 \textbf{as} RL = 4
\State \textbf{transfor} BPM to ML:
\State \qquad 70 to 90 \textbf{as} ML = 1
\State \qquad 90 to 110 \textbf{as} ML = 2
\State \qquad 110 to 130 \textbf{as} ML = 3
\State \textbf{run} Kendall rank using ML:
\State \qquad \textbf{obtain} correlation coefficient \textbf{as} cc
\State \textbf{define} game type \textbf{as} gt
\State \textbf{synthesize} cc, gt \textbf{vs.} RL \textbf{for} conclusion
\end{algorithmic}
\end{algorithm}

Undoubtedly, music that better matches the scene will increase the player's immersion, thereby affecting the player's time perception. However, in reality, a gamer who often play Japanese RPG game will not suddenly try a western-style RPG game which is in a completely different style. In other words, participant A’s data in our experiment has no practical meaning. Moreover, the matching degree cannot be perfectly quantified. In the experiment, we selected two sets of music with totally different styles, but in the practical situation, this parameter will largely depend on the participant’s personal preference when the selected music is in a similar style.

Since the matching degree has no obvious effect on the time perception, what is the reason for the huge difference in the user's estimation of time in the four scenarios? Our guess was how much the participants liked the music and the setting, and how familiar they were with it.

\begin{figure}[h]
\centering
\includegraphics[width=0.8\linewidth]{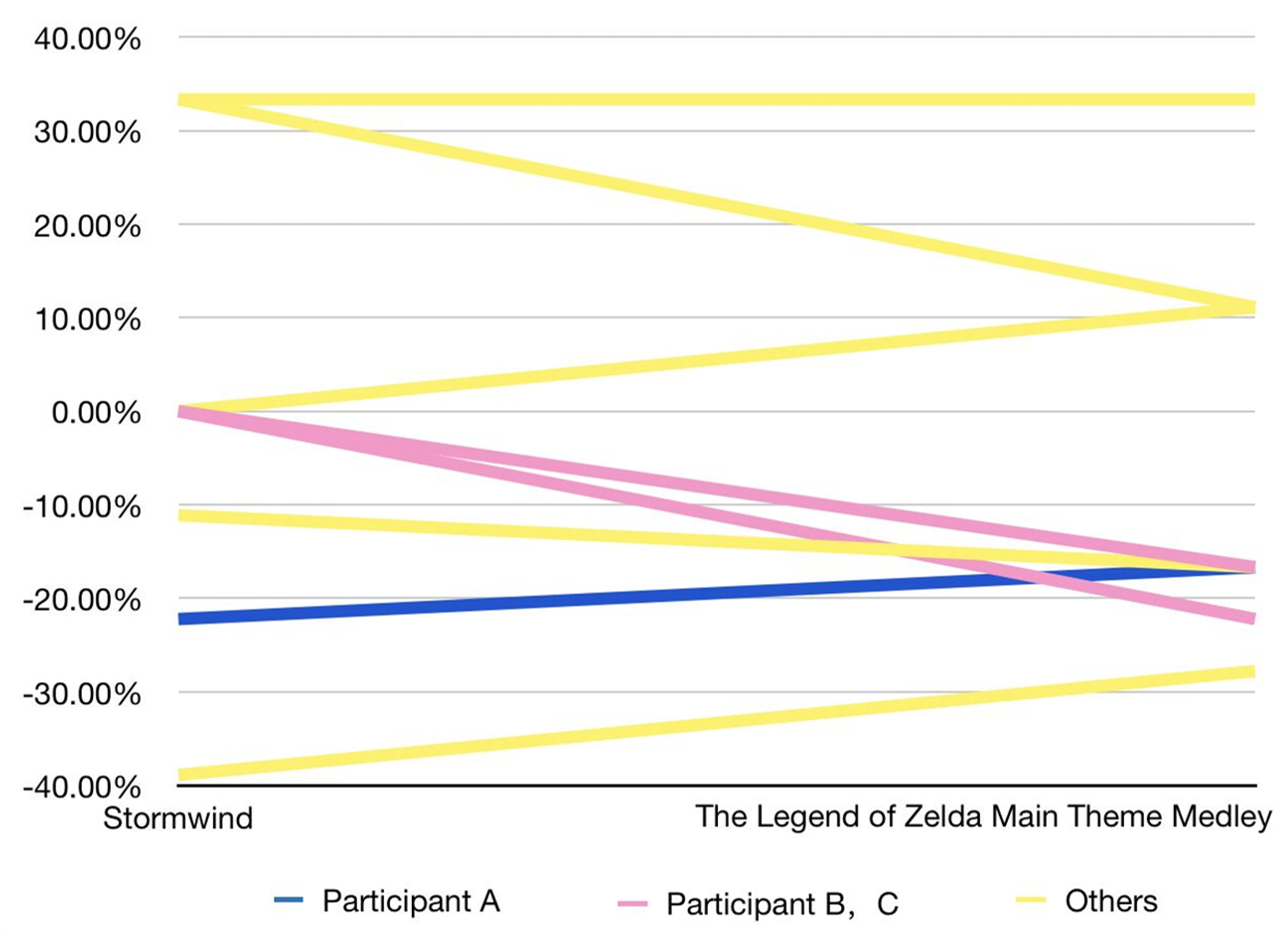}
\caption{The Influence of Different Music on Time Estimation in One Scene}
\label{fig:fig7}
\end{figure}

Finally, in response to this result, we conducted a more in-depth investigation and analysis. We collected all types of games that the participants usually play(last step in Algorithm 2), and their satisfaction with the test music, and found that the experimental results are closely related to participants’ gaming experience. The \autoref{fig:fig7} lists the results of the two sets of experimental music in this scene. For participants A, he only plays Japanese RPG games, hence he believes Legend of Zelda is more compatible with the scene, and he shows a longer time perception in this music. Participants B and C who mainly play western-style RPG games show the opposite result. They think Stormwind is more suitable for the scene with faster time perception. For the rest of the participants, they do not play RPG games, the degree of matching depends on whether the testing music itself is songful or not.

Additionally, \autoref{tab:tab5} shows the data of two experimental groups in a specific scene . The reason why the control group is not added is that the subjects generally have no special feeling for the test music of the control group, and there is no significant difference in likeability and familiarity. Under the overall picture, it can also be seen that the matching degree between scenes and music does not significantly affect time perception. This is because everyone has a different definition of a matching degree. As can be seen from the table, almost all participants' views on the matching degree of the two pieces of test music are contrary to what we assumed, although we have selected test music and the original soundtrack with extremely different styles.

\begin{table}[h]
\centering
\caption{Influence of Matching Degree(1-5), Likeability(1-5) and Familiarity(0 or 1) on Time Estimation(bolded: Experimental Group 1, unbolded: Experimental Group 2)}
\label{tab:tab5}
\scalebox{0.8}{\begin{tabular}{ccccccccc} 
\toprule
\multirow{2}{*}{Scene}         & \multirow{2}{*}{Number} & \multicolumn{2}{c}{Familiarity} & \multicolumn{2}{c}{Likeability} & \multicolumn{2}{c}{Matching Degree}          & \multicolumn{1}{c}{\multirow{2}{*}{TD}}  \\ 
\cline{3-8}
                               &                         & scene              & Music                       & scene              & Music      & Theoretical & \multicolumn{1}{l}{Subjective} & \multicolumn{1}{c}{}                     \\ 
\cline{1-9}
\multirow{16}{*}{Stormwind}    & \multirow{2}{*}{\#1}    & \multirow{2}{*}{1} & \multirow{2}{*}{1}          & \multirow{2}{*}{5} & \textbf{3} & \textbf{5}  & \textbf{3}                     & \textbf{-44\%}                           \\
                               &                         &                    &                             &                    & 4          & 1           & 4                              & -22\%                                    \\
                               & \multirow{2}{*}{\#2}    & \multirow{2}{*}{1} & \multirow{2}{*}{0}          & \multirow{2}{*}{4} & \textbf{3} & \textbf{5}  & \textbf{4}                     & \textbf{-11\%}                           \\
                               &                         &                    &                             &                    & 4          & 1           & 4                              & -17\%                                    \\
                               & \multirow{2}{*}{\#3}    & \multirow{2}{*}{1} & \multirow{2}{*}{1}          & \multirow{2}{*}{5} & \textbf{5} & \textbf{5}  & \textbf{5}                     & \textbf{-39\%}                           \\
                               &                         &                    &                             &                    & 5          & 1           & 5                              & -28\%                                    \\
                               & \multirow{2}{*}{\#4}    & \multirow{2}{*}{1} & \multirow{2}{*}{1}          & \multirow{2}{*}{5} & \textbf{4} & \textbf{5}  & \textbf{4}                     & \textbf{0\%}                             \\
                               &                         &                    &                             &                    & 4          & 1           & 5                              & +11\%                                    \\
                               & \multirow{2}{*}{\#5}    & \multirow{2}{*}{1} & \multirow{2}{*}{1}          & \multirow{2}{*}{5} & \textbf{3} & \textbf{5}  & \textbf{3}                     & \textbf{+33\%}                           \\
                               &                         &                    &                             &                    & 4          & 1           & 4                              & +11\%                                    \\
                               & \multirow{2}{*}{\#6}    & \multirow{2}{*}{1} & \multirow{2}{*}{0}          & \multirow{2}{*}{4} & \textbf{3} & \textbf{5}  & \textbf{4}                     & \textbf{0\%}                             \\
                               &                         &                    &                             &                    & 5          & 1           & 5                              & -22\%                                    \\
                               & \multirow{2}{*}{\#7}    & \multirow{2}{*}{0} & \multirow{2}{*}{0}          & \multirow{2}{*}{5} & \textbf{2} & \textbf{5}  & \textbf{3}                     & \textbf{+33\%}                           \\
                               &                         &                    &                             &                    & 4          & 1           & 4                              & +11\%                                    \\
                               & \multirow{2}{*}{\#8}    & \multirow{2}{*}{1} & \multirow{2}{*}{1}          & \multirow{2}{*}{5} & \textbf{4} & \textbf{5}  & \textbf{3}                     & \textbf{0\%}                             \\
                               &                         &                    &                             &                    & 5          & 1           & 5                              & -17\%                                    \\ 
\midrule
\multirow{2}{*}{\textbf{Mean}} &                         &                    &                             &                    &            &             &                                & \textbf{-3.5\%}                          \\
                               &                         &                    &                             &                    &            &             &                                & -9.1\%                                   \\
\bottomrule

\toprule
\multirow{2}{*}{Scene}         & \multirow{2}{*}{Number} & \multicolumn{2}{c}{Familiarity}         & \multicolumn{2}{c}{Likeability} & \multicolumn{2}{c}{Matching Degree}          & \multirow{2}{*}{TD}  \\\cline{3-8}
                               &                         & scene              & Music              & scene              & Music      & Theoretical & \multicolumn{1}{l}{Subjective} &                      \\ 
\cline{1-9}
\multirow{16}{*}{Scenery}      & \multirow{2}{*}{\#1}    & \multirow{2}{*}{1} & \multirow{2}{*}{0} & \multirow{2}{*}{3} & \textbf{3} & \textbf{4}  & \textbf{3}                     & \textbf{-44\%}       \\
                               &                         &                    &                    &                    & 4          & 4           & 2                              & -44\%                \\
                               & \multirow{2}{*}{\#2}    & \multirow{2}{*}{0} & \multirow{2}{*}{0} & \multirow{2}{*}{3} & \textbf{4} & \textbf{4}  & \textbf{5}                     & \textbf{-11\%}       \\
                               &                         &                    &                    &                    & 3          & 4           & 3                              & -17\%                \\
                               & \multirow{2}{*}{\#3}    & \multirow{2}{*}{0} & \multirow{2}{*}{0} & \multirow{2}{*}{5} & \textbf{5} & \textbf{4}  & \textbf{5}                     & \textbf{-39\%}       \\
                               &                         &                    &                    &                    & 3          & 4           & 4                              & -28\%                \\
                               & \multirow{2}{*}{\#4}    & \multirow{2}{*}{0} & \multirow{2}{*}{0} & \multirow{2}{*}{4} & \textbf{4} & \textbf{4}  & \textbf{3}                     & \textbf{+22\%}       \\
                               &                         &                    &                    &                    & 4          & 4           & 3                              & +11\%                \\
                               & \multirow{2}{*}{\#5}    & \multirow{2}{*}{1} & \multirow{2}{*}{1} & \multirow{2}{*}{5} & \textbf{4} & \textbf{4}  & \textbf{4}                     & \textbf{-11\%}       \\
                               &                         &                    &                    &                    & 5          & 4           & 4                              & -33\%                \\
                               & \multirow{2}{*}{\#6}    & \multirow{2}{*}{0} & \multirow{2}{*}{0} & \multirow{2}{*}{4} & \textbf{4} & \textbf{4}  & \textbf{5}                     & \textbf{-22\%}       \\
                               &                         &                    &                    &                    & 5          & 4           & 2                              & -33\%                \\
                               & \multirow{2}{*}{\#7}    & \multirow{2}{*}{0} & \multirow{2}{*}{1} & \multirow{2}{*}{5} & \textbf{4} & \textbf{4}  & \textbf{4}                     & \textbf{+33\%}       \\
                               &                         &                    &                    &                    & 3          & 4           & 3                              & +33\%                \\
                               & \multirow{2}{*}{\#8}    & \multirow{2}{*}{0} & \multirow{2}{*}{0} & \multirow{2}{*}{5} & \textbf{5} & \textbf{4}  & \textbf{5}                     & \textbf{0\%}         \\
                               &                         &                    &                    &                    & 4          & 4           & 5                              & +11\%                \\ 
\midrule
\multirow{2}{*}{\textbf{Mean}} &                         &                    &                    &                    &            &             &                                & \textbf{-9.0\%}      \\
                               &                         &                    &                    &                    &            &             &                                & -12.5\%              \\
\bottomrule
\end{tabular}}
\end{table}

However, based on the test data of the Stormwind scene, it is the user's likeability and familiarity with the scene, music, and other elements that have the greatest impact on time perception. This point is well reflected in the data of No. 3. This user plays a lot of games in daily life, and most of them are large-scale games with grand scenes which is quite similar to Stormwind. Because of this, full marks of likeability and familiarity were given in the questionnaire, and the test results were much shorter than the pre-set time. On the contrary, the participant No.7, who had never been exposed to this type of scene but was very familiar with JRPG (Japanese Role-Play Game) which is group 2 music comes from, showed a sharp contrast between the two test results. In other words, familiarity with the scene and music helped shorten subjective time estimates.

Furthermore, due to the limited selection of scenes in Stormwind, it is difficult to tell whether the likeability of music also affects time perception. To confirm this, let's look at data from another scene. It is a neutral scene that will not be universally objectionable or attractive, which can maximize the influence of music-likeability on time perception. We selected two kinds of light music with different styles (\autoref{tab:tab3}), and asked the subjects to choose the one they preferred by comparison when filling in the questionnaire. According to the data in the table, except for No. 7 and No. 4, all the results prove that the music users prefer helps to shorten the subjective time estimate.

In summary, background music can not be counted as strong Zeigebers in a VR environment, either from the aspect of pace or matching degree. As an art form, music cannot be fully quantified. The same music can have a completely different listening experience by changing a few timbre and rhythm. At the same time, we cannot ignore the fact that background music lacks musicality to some extent, and it generally plays the role of the setting of the atmosphere\cite{Henry2017TheDB}. It is not feasible for VR environment designers to focus on background music in the expectation of having a significant impact on the player's time perception since it can never have a direct effect like game content or game screen. Thinking about designing a piece of background music that is as familiar and resonant as possible to the target users will be worked.

As an auxiliary way, background music can never be mighty influential as the game content and game graphics. In other words, it cannot play a decisive role. However, as an auxiliary means to regulate time perception in a virtual reality environment, music can influence users' time perception by affecting immersion to some extent. On the other hand, if individual differences can be effectively quantified, perhaps the effect of music can be further subdivided. Therefore, we take attention as one of the important manifestations of individual differences and add it to the next part.

\subsection{Acceleration Effect of Low Cognitive Load}
\autoref{tab:tab6} synthesizes original data. Through a posterior analysis, we can simply calculate Cohen's d is 3.4891, effect size (r) is 0.8675, the sample size is 36. Then we set $\alpha$ is 0.05, using these parameters to calculate statistical test force, which is 0.8174, and it is in line with the expectation.

\begin{sidewaystable}
\sidewaystablefn%
\begin{center}
\caption{Cognitive load and Attention in Pathfinding Task(CL:Cognitive Load)}
\label{tab:tab6}
\centering
\scalebox{0.7}{
\begin{tabular}{cccccccccccccccc}
\toprule
      && \textbf{\#1} & \textbf{\#2} & \textbf{\#3} & \textbf{\#4} & \textbf{\#5} & \textbf{\#6} & \textbf{\#7} & \textbf{\#8} & \textbf{\#9} & \textbf{\#10} & \textbf{\#11} & \textbf{\#12}& \textbf{Mean}& \textbf{Variance} \\ \midrule
     & TD & -20\%      & 37.50\%    & -14.29\%   & -27.28\%   & 62.50\%    & -25\%      & -25\%      & -10\%      & -11.11\%   & -36.17\%    & -5.66\%     & -38.46\%  & -0.09\% & 0.09  \\
High CL  & Noise/data & 0/29       & 0/50       & 0/27       & 0/37       & 0/67       & 0/29       & 0/37       & 0/58       & 0/32       & 0/70        & 0/63        & 0/37    &&    \\
(Text) & Attention level & 2.86       & 3.58       & 4.59       & 4.78       & 4.88       & 3.97       & 5.92       & 5.90       & 4.52       & 3.89        & 4.95        & 4.24     &4.51&0.79   \\
     & Attention   & 35.44      & 44.26      & 58.22      & 61.86      & 62.57      & 50.48      & 76.11      & 76.16      & 57.87      & 49.23       & 64.34       & 53.81    &&   \\ \midrule
     & TD & -11.11\%   & 0          & -27.28\%   & -33.33\%   & -7.70\%    & 0          & 20\%       & 42.86\%    & -16.67\%   & -24.53\%    & -9.10\%     & 0      & -0.06\% & 0.04     \\
Medium CL  & Noise/data & 0/69       & 0/73       & 0/54       & 0/35       & 0/50       & 0/35       & 0/47       & 0/27       & 1/53       & 0/38        & 0/90        & 0/44    &&    \\

(Graphic) & Attention level & 3.75       & 4.06       & 3.68       & 4.63       & 3.18       & 3.71       & 4.53       & 4.29       & 4.17       & 2.95        & 4.21        & 5.02    &4.02&0.35    \\
     & Attention   & 44.79      & 51.69      & 46.54      & 59.34      & 39.08      & 47.29      & 52.28      & 53.85      & 54.23      & 36.51       & 53.53       & 64.51    &&   \\ \midrule
     & TD & -40\%      & 0          & -57.14\%   & -60\%      & -50\%      & -50\%      & -50\%      & -28.57\%   & -14.89\%   & -50\%       & -75.61\%    & -20.53\%  & -0.41\% & 0.05  \\
Low CL  & Noise/data & 1/46       & 0/54       & 0/39       & 0/45       & 0/59       & 0/31       & 0/52       & 0/37       & 0/74       & 0/37        & 2/37        & 3/66    &&    \\
(Video) & Attention level & 2.76       & 3.74       & 4.35       & 4.60       & 4.67       & 4.58       & 5.17       & 5.59       & 2.99       & 5.30        & 2.47        & 4.03     &4.19&1.04   \\
     & Attention   & 33.54      & 47.50      & 55.39      & 59.69      & 59.49      & 59.48      & 67.73      & 72.81      & 37.66      & 67.46       & 30.25       & 51.03    &&   \\ \bottomrule
\end{tabular}}
\end{center}
\end{sidewaystable}

\begin{figure}[h]
\centering
\includegraphics[width=0.8\linewidth]{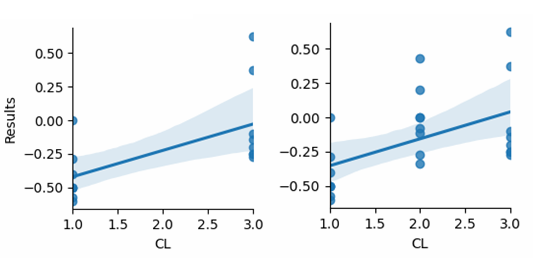}
\caption{Prediction line between CL and TD with (right) and without (left) graphic data regardless of Attention (confidence level 95\%)}
\label{fig:fig8}
\end{figure}

Overall data can support the hypothesis, but only part of the data, mainly graphic data, is not completely consistent. From the questionnaire and testing process, we found for some participants, the graphic guidance was not even as explicit as the text guidance. Someone with a weak ability to analyze picture elements, text guidance even had a lower cognitive load. That is the reason why graphic data is not completely consistent. As can be seen from the comparison of \autoref{fig:fig8}, the overall predicted trend remains unchanged with or without graphic data. The data within the confidence interval is from ordinary players, and outside are from top players. According to the questionnaire survey, this is because these participants learned game mechanisms quickly and felt the task was too simple, thus their actual cognitive load was much lower than the ordinary player, leading to their faster subjective time estimation. 

The difference of graphic data lies in the correlation coefficient which is 0.51 when graphic data is present and becomes 0.61 when removed. This suggests that cognitive load is a low-intensity correlation zeitgeber without considering graphic data, which can play a certain role in time judgment. However, it can only be regarded as a moderately correlated factor if considering graphic data. Meanwhile, we use the evaluation function SCORE to determine the quality of the regression equation. The best score of the SCORE function is 1.0, and a negative value may occur in the case of a poor model. With and without graphic data, the scores were 0.43 and 0.59, respectively. Therefore, the disadvantages of graphic data in this analysis can be further demonstrated.

\begin{figure*}[h]
\centering
\includegraphics[width=\linewidth]{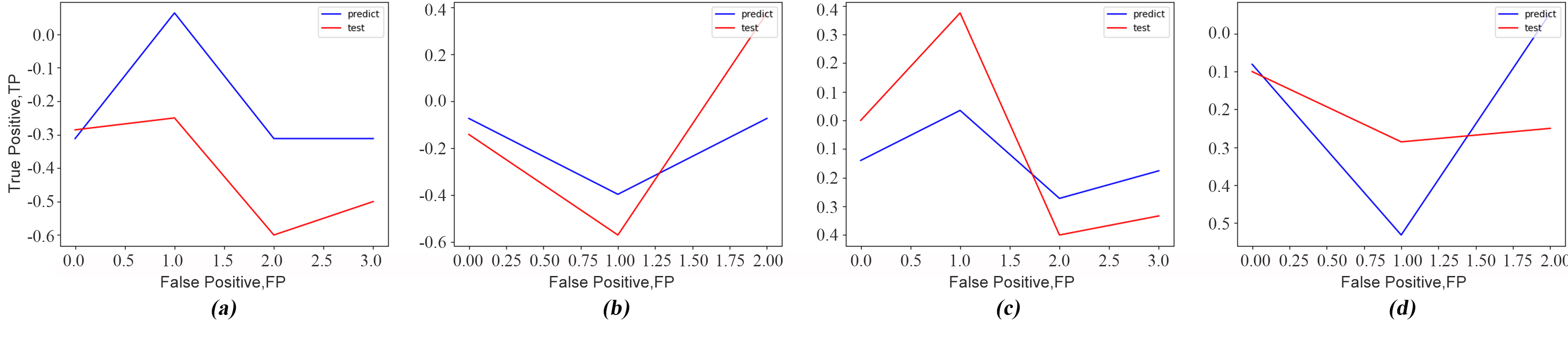}
\caption{ROC curve exclude Attention of predicted value and actual value with(a) and without graphic data (b), and include Attention predicted value and actual value with(c) and without attention (d)}
\label{fig:fig9}
\end{figure*}

The analysis of the Receiver Operating Characteristic Curve (ROC) gives us the same conclusion in \autoref{fig:fig9} (a) and (b). Specifically, which is based on a series of different ways of binary classification (boundary value or decision threshold), with true positive rate as the ordinate and abscissa to draw the curve of the rate of false positives\cite{10.1093/clinchem/39.4.561}. The traditional diagnostic test evaluation methods have a common feature that the test results must be divided into two categories and then analyzed statistically. ROC curve evaluation method is different from the traditional ones, need not this limit\cite{10.1093/clinchem/39.4.561}. In conclusion, the ROC curve evaluation method applies to a wider range, which is also the reason why this method is used in this paper.

In general, our method of analyzing relevant data is as shown in algorithm 3. Another dimension data (attention) will be added later, but the analysis process remains the same.

\begin{algorithm}[h]
\caption{CL $\&$ Attention Data Analysis Logic}
\label{alog:a3}
\begin{algorithmic}[1]
\State \textbf{run} priori test \textbf{to obtain} sample size
\State \textbf{cal} conhen’d, effect size:
\State \qquad \textbf{run} post hoc \textbf{to verify} data set
\State \textbf{try} Lagrange Interpolation \textbf{to replace} outliers
\State \textbf{cal} MLR and \textbf{verify} by:
\State \qquad \textbf{draw} prediction curve
\State \qquad \textbf{draw} ROC:
\State \qquad \qquad \textbf{cal} fitting degree(Amplitude $\&$ Tendency)
\State \qquad \textbf{run} SCORE
\State \qquad \textbf{cal} correlation coefficient
\State \textbf{to obtain} predictive equation
\end{algorithmic}
\end{algorithm}

\subsection{Embodiment of Individual: Attention}
We added data collected by a portable EEG device TGAM to calculate attention, which reflected how focus the participants were. Particularly, an optimization algorithm was used to improve EEG data\cite{2020Attention,2020The}, and we generate two parameters: one is attention whose value varying in the range of 0-100, and the other is attention level divided into eight grades (0-7). 

Specific results are shown in \autoref{tab:tab4}, where noise/data represents the proportion of noise interference in the collected EEG data, which can reflect that our EEG data has little noise and good quality. As early as 1982 with the introduction of the EEG equipment analysis of zeitgeber precedent \cite{doi:10.3109/00207458309148662},  the experiment explored the effects of sleep on human circadian rhythm using volunteer’s mental health evaluated by EEG data. In our experiment, we only focus on whether attention contributed to time estimation.

\begin{gather}
\label{eq:CL_no_Attention}
        TD=\alpha+\gamma\times \mbox{CL}
\end{gather}

When Attention is not considered, the final fitting linear regression equations is as \autoref{eq:CL_no_Attention}, $\alpha$ and $\gamma$ equal to -0.63 and 0.21 when including graphic data, while it becomes -0.50 and 0.16 when excluding. For the equation with graphic data, we tried to replace outliers with Lagrange interpolation. However, due to the relatively small data set, the effect has not been significantly improved. The CL(Cognitive Load) is set as 3 for text guidance (high Load), 2 for graphic guidance (medium Load), and 1 for video guidance (low Load). Due to the length of our experiment time, this prediction equation is more accurate in short-term scenes and has certain reference significance for long-term scenes. However, due to insufficient data sets and the lack of consideration of individual differences, the prediction results are not satisfactory.

It can be seen from the above results that lower cognitive load leads to shorter time estimates. However, some data, especially graphic data, do not strongly support this conclusion.The main reason, as in the music experiment, is the influence of individual differences. An interesting question can be proposed: time estimates can also be influenced by individuals' differing understandings of the interesting, complexity, difficult and other attributes of testing task\cite{FINK20011009}. One important parameter that we quantify individual difference is Attention.

\begin{table}[h]
\caption{Comparison of Fitting Results in Four Cases}
\centering
\scalebox{0.85}{
\begin{tabular}[t]{cccccccc}
\toprule
           & \multirow{2}*{Graphic Data}                 & \multirow{2}*{Attention}                  & \multirow{2}*{Score}              & Correlation coefficient  & \multicolumn{2}{c}{ROC Fitting Degree} & \multirow{2}*{p-value}  \\ \cmidrule{6-7} 
           & \multicolumn{1}{l}{} & \multicolumn{1}{l}{} & \multicolumn{1}{l}{} & (TD vs. CL) & Amplitude            & Tendency           \\ \midrule
\textbf{\#1} & include                     & exclude                   & 0.43          & 0.51     & 68\%          & 65\%      &0.17   \\
\textbf{\#2} & exclude                    & exclude                   &   0.59         & 0.61     & 64\%          & 88\%     &0.74  \\
\textbf{\#3} & include                     & include                  & 0.54          & 0.51     & 86\%          & 91\%      &0.09   \\
\textbf{\#4} & exclude                    & include                  &    0.40        & 0.61     & 58\%          & 43\%   &0.06   \\ \bottomrule
\end{tabular}}
\label{tab:tab7}
\end{table}

After the Attention analysis is added, we get four models as shown in \autoref{tab:tab7}. In order to prove that the prediction can be more accurate after we add the Attention analysis, we compare the prediction results with or without graphic data. This is because, through the questionnaire survey and the previous analysis, the graphic data are the data that can reflect the individual differences. As can be seen from the data in \autoref{tab:tab7}, the Score is 0.40 and 0.53 respectively when there is or is not graphic data. ROC curve (\autoref{fig:fig9} (c) and (d)) also gives similar results, the four groups of data in \autoref{tab:tab1} is corresponding to (a) to (d) in \autoref{fig:fig2} respectively. Obviously, image prediction tendency with graphic data and Attention is more consistent with the real situation, and its fitting degree is higher in both amplitude and tendency.

Finally, the prediction equation with Attention as \autoref{eq:CL_with_attention}. Specifically, $\beta$, $\theta$ and $\lambda$ are equal to -0.33, 0.17 and 0.04 respectively. Adding Attention as a parameter can greatly improve the prediction accuracy with graphic data, and to some extent, it is even better than eliminating the graphic data directly.

\begin{equation}
\label{eq:CL_with_attention}
        TD=\beta+\theta\times \mbox{CL}-\lambda\times \mbox{Attention\_level}
\end{equation}

Moreover, an interesting phenomenon can be found from the attention data: the data that cannot prove the positive correlation between cognitive load and subjective time estimation, such as the graphic data of subjects 6, 7, and 8, all have significantly lower attention in horizontal comparison(\autoref{tab:tab6}). Furthermore, Due to the task design of our experiment, participants who were top players or showed low interest in our experiment had low attention performance (e.g. 1, 2). However, participants who showed high attention, longer time estimation did not fully follow the rule of proportional to cognitive load (e.g. 6, 7, 8). If we made the task harder, probably the effect of attention would be even more pronounced. This aspect also shows that we must add attention to data analysis, which quantifies individual differences to a certain extent. There is a connection between attention and time: in general, higher attention leads to a shorter time estimation.

\section{Conclusion}
After the above experiment and analysis, we can draw the following conclusions:

\begin{itemize}
\item As a short-term zeitgeber, highly sensitive colors are conducive to enhancing the instantaneous reaction of users and shortening the estimation of time (CadetBlue > Chartreuse $\geq$ Yellow > Grey). 
\item In the VR environment, as an auxiliary method to control time perception music, is mainly realized by the influence of immersion. The more familiar and beloved the music is, the more conducive it is to shorten the time perception.
\item Cognitive load and attention are two vital Zeitgebers that affect time perception in VR together. Low cognitive load and high attention speed up time perception the most.
\item Attention reflects individual differences to some extent.
\end{itemize}

Zeitgebers have some different approaches to influence time perception in the physical and virtual world(\autoref{fig:fig2}). In the VR environment, reasonable use of the strong correlation between cognitive load and attention on time perception can optimize the design of scenes. Zeitgebers such as color and music can also be used as auxiliary methods to regulate time perception. These may have implications for the physical world as well.

However, due to the influence of factors as screen resolution and scene optimization, the principle of time perception in VR is different. Learning the same content, a person in the VR environment will spend less time than a person in the physical world, also the person in the physical world will feel time flows slower. \cite{MAKRANSKY2019225}. Besides, there is an interaction between immersion and attention in the VR environment, that is, better immersion leads to higher attention, which further affects subjective time estimation\cite{8585373}. Therefore, it can be said that the VR environment itself has already affected the user's time estimation.

\section{Limitation \& Future Work}
There is no doubt that there is more work to be done. The emotion, which is a vital zeitgeber that can reflect the specific state of the user in the virtual environment and play a decisive role in the time perception is not taken into consideration in this study. Particularly, studies have shown that people who can regulate emotions can reduce or even ignore the impact of emotions on time perception\cite{2020PotentialInteraction}. However, there are too many kinds of emotions that may play a role in time perception, and it is difficult to find an effective way to quantify emotions due to the limitations of equipment and the experimental environment. Thus, the emotion is factor emotion is not considered in this study and it needs more work in the future. 

In addition, our researches on color have type limitations, the selection of experiment colors is restricted by the color brightness. Also, the experiment method on Zeitgeber music needs to be improved as our data analysis cannot strongly prove that music has a great impact on time perception. This is because it is hard to quantify individual differences and multiple experiments were conducted in a short time causing the participant familiar more with the following experiment and leading to the experiment result variance. To minimize such effect, for the same Zeitgeber, every experiment was separated by at least 2 days. However, this experiment method still needs to be improved as the time estimation will be inevitably shorter when the participant is more familiar with the environment. Furthermore, a higher ecological validity test environment is necessary. If we can design a highly modular virtual reality environment, makes the environment variables or factors can be artificially adjust, rather than as used in this article some existing VR environment, presumably can improve the rigor and reliability of the experiment.

As for the limitations, first of all, we want to improve the influence of emotions on the sense of time. In the physical world, emotion is also a very important Zeitgeber like attention\cite{YAMADA20111835,10.3389/fnint.2011.00033}. In the experiments in this article, we use 1 and 0 to represent the presence or absence of tension is the simplest way to quantify emotions, but the effect is not ideal due to various reasons. Finding a better way to quantify this parameter is valuable.

Secondly, as an important parameter,  the level of immersion is vital to determine the quality of a VR environment whether the user can truly immerse themselves in the virtual world. Besides, immersion can also be regarded as a zeitgeber, since certain immersive experiences can make people feel time flows faster. Although there have been some studies that can refer to\cite{10.1162/105474698565686,article2}, there is no universal method that can effectively parameterize users’ immersion now. Indeed, the difficulty of expressing subjective feelings with objective parameters is not only reflected in the hardware conditions but also in data analysis. In particular, immersion can be subdivided into scene immersion and emotional immersion\cite{7965655}, it is very difficult to achieve them at the same time. In that case, we will also focus on creating a simple, efficient, and low-cost method to quantify immersion.


\appendix

\section{Video Game References}
(i)   Half-Life: Alyx(2020) Developed by Valve.\\
(ii)  Minecraft(2009) Developed by Mojang Studios.\\
(iii) World of Warcraft(2004) Developed by Blizzard Entertainment.\\
(iv) The Legend of Zelda: Breath of the Wild(2017) Developed by Nintendo.

\section{Hardware Setup}
Our IVEs ilnclude few parts. Firstly, as illustrated in Fig. 1, participants wear an HTC VIVE Pro HMD to experiment with the immersive environment. The HMD provides a resolution of 1440 $\times$ 1600 pixels per eye, and the binocular resolution is 2880 $\times$ 1600 with a refresh rate of 90 Hz and an approximately 110 degrees diagonal field of view. Secondly, we track its position and orientation with SteamVR$^{TM}$ tracking system. Finally, we use Neurosky TGAM(Thinkgear Asic Module) to collect EEG data. Non-immersive display setup used a monoscopic display without head-tracking, which presented on a Acer XV272UP display with a resolution of 2560 $\times$ 1440 and a refresh rate of 144Hz. We use a workstation for rendering and as the experimental platform, whose processor is 4.90 GHz 8 Core i7-9700K processor, 32 GB(3200Mhz) of main memory, and a GeForce GTX 2070 8GB graphics card.


\bibliography{sn-bibliography}


\end{document}